\documentclass[%
superscriptaddress,
 reprint,
 amsmath,amssymb,
 aps,
]{revtex4-2}

\usepackage{graphicx}
\usepackage{dcolumn}
\usepackage{bm}

\usepackage[colorlinks,linkcolor=blue,citecolor=blue,urlcolor=blue]{hyperref}
\usepackage{ulem}

\def\bs#1{\boldsymbol{#1}}

\begin{document}


\title{
Anomalous Circularly Polarized Light Emission induced by the Optical Berry Curvature Dipole
}

\author{Yizhou \surname{Liu}}
\affiliation{School of Physics Science and Engineering, Tongji University, Shanghai 200092, China}
\affiliation{Department of Condensed Matter Physics, Weizmann Institute of Science, Rehovot 76100, Israel}
\author{Binghai \surname{Yan}}
\email{binghai.yan@weizmann.ac.il}
\affiliation{Department of Condensed Matter Physics, Weizmann Institute of Science, Rehovot 76100, Israel}

\date{\today}

\begin{abstract}
The ability to selectively excite light with fixed  handedness is crucial for circularly polarized light emission. It is commonly believed that the luminescent material chirality determines the emitted light handedness, regardless of the light emitting direction. In this work, we propose an anomalous circular polarized light emission (ACPLE) whose handedness actually relies on the emission direction and current direction in electroluminescence. In a solid semiconductor, the ACPLE originates in the band structure topology characterized by the optical Berry curvature dipole. ACPLE exists in inversion-symmetry breaking materials including chiral materials. We exemplify the ACPLE by estimating the high circular polarization ratio in monolayer WS$_2$. In addition, the ACPLE can be further generalized to magnetic semiconductors in which the optical Berry curvature plays a leading role instead. Our finding reveals intriguing consequences of band topology in light emission and promises optoelectric applications. 
\end{abstract}

\maketitle

\section{Introduction}

Circularly polarized (CP) light emission is important for various advanced technologies including 3D display, optoelectrics, information storage, and quantum spintronics \cite{Zhang2020review, Han2018review}. The CP light can be generated through optical fibers composed by a linear polarizer followed by a quarter-wave plate or CP light emission (CPLE) processes including photoluminescence (PL) and electroluminescence (EL) in chiral organic luminophores \cite{Riehl1986,Sanchez2015,Longhi2016,Sang2020,Deng2021,Crassous2023} and chiral hybrid organic–inorganic perovskites \cite{Kim2021,Long2020}. It is commonly believed that the favored light handedness in both PL and EL are determined by the material chirality but independent of the emitting direction. Recently, a distinct CP EL was reported in the chiral polymers where the favored light handedness depends on both the emitting direction and polymer chirality \cite{Wan2023NatPhoton} . Furthermore, the emission-direction-sensitive CPLE generates giant net polarization ratio of CPLE by orders of magnitude. 
The anomalous CP light emission (ACPLE) was interpreted by the orbital-momentum locking, a topological nature of the chiral molecule electronic structure\cite{Liu2021,Wan2023NatPhoton}. 

Compared to chiral organic polymers, we are inspired to explore the ACPLE in crystalline solids where the band structure topology \cite{Hasan2010, Qi2011, Yan2017, Armitage2018} can play a significant role. For example, the Berry curvature and Berry connection\cite{Xiao2010review} is extensively considered in the content of light absorbance and the resultant photocurrent, such as the photogalvanic effects \cite{vonBaltz1981,Sipe2000, Hosur2011, Morimoto2016, Holder2020,deJuan2017}, nonlinear anomalous Hall effect \cite{Sodemann2015,Zhang2018,Ma2017,kang2019nonlinear} and the valley Hall effect \cite{Xiao2007,Yao2008}.

In solids, giant CP EL was indeed observed in inversion-breaking semiconductors, WSe$_2$ \cite{Zhang2014} and WS$_2$ \cite{Pu2021WS2_exp} monolayers, where emitted light handedness is switchable by reversing the current flow. It was rationalized by the valley polarization and asymmetric band structures of electrons and holes. Additionally, a magnetization switched CP PL was also reported in a magnetic insulator CrI$_3$ \cite {Seyler2018}, which was interpreted by local molecular orbital transitions of Cr$^{3+}$ cations. From the theoretical aspect, the electronic wavefunctions and topology should play an essential role there but remain largely unexplored, although the band dispersion (\textit{e.g.} band shift, and triangular warping etc. \cite{Wang2020}) was recently discussed in this content.

In this work, we formulate the ACPLE that exists in inversion-breaking solids (including chiral solids) by the optical Berry curvature dipole. Different from the ordinary CP EL, the favored light handedness depends both on the direction of applied current and the light-emitting direction. Here, ACPLE can be regarded as a reverse process of the circular photogalvanic effect where CP light irradiation generates a dc current \cite{Sipe2000}. We demonstrate the ACPLE in the strained WS$_2$ monolayer. The CP EL in WS$_2$ shows handedness switching by the direction of applied strain, current flow, and emitting direction. In addition, the ACPLE can also appear in magnetic solids, which originate in the optical Berry curvature. Our theory rationalizes the current direction or magnetization-switchable light handedness in previous experiments based on CrI$_3$ or WS$_2$. Our predictions can be further verified by detecting the emission-direction dependent handedness in present devices. The ACPLE promises applications in exotic CP light-emitting diodes and ultrafast optoelectronic information applications.

\begin{figure*}
    \centering
    \includegraphics[width=\linewidth]{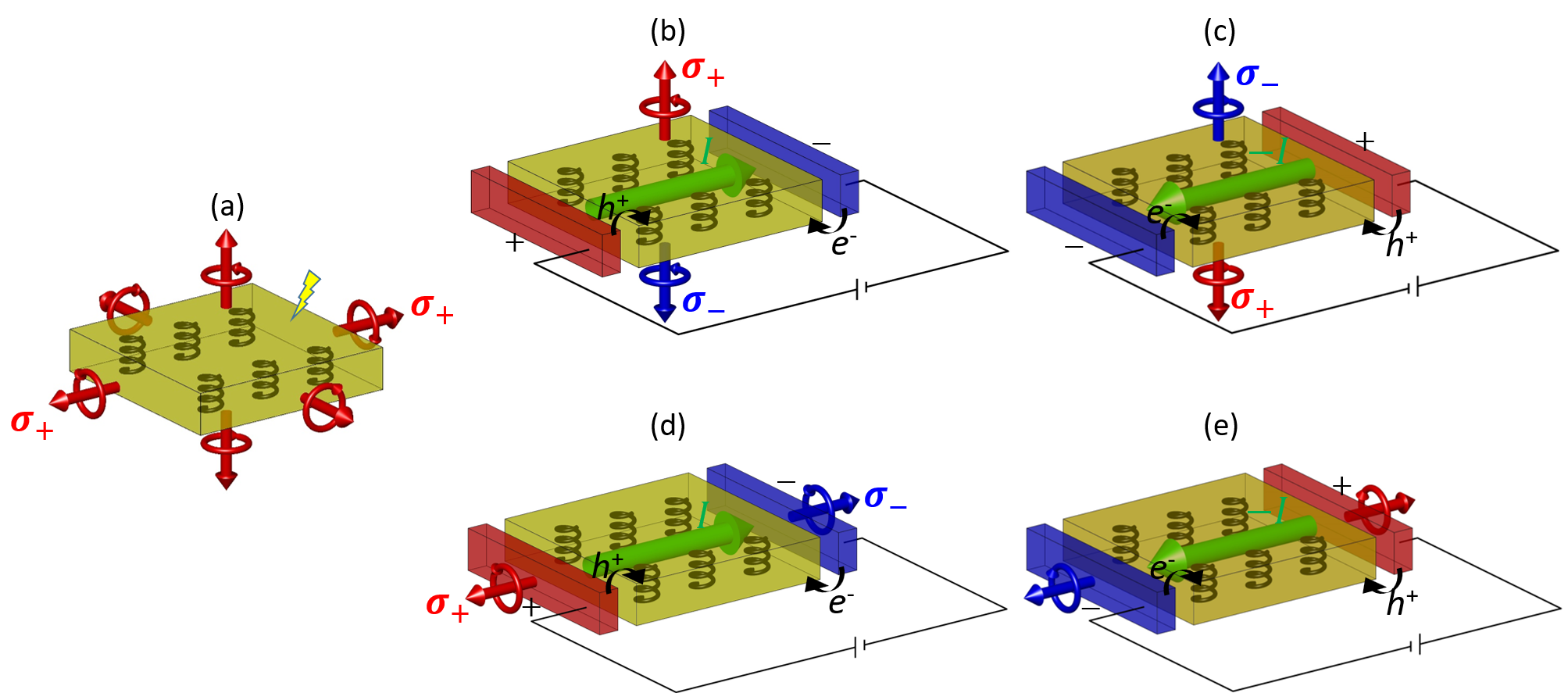}
    \caption{
    Schematics of the (a) ordinary CPLE and (b)-(e) ACPLE, respectively. (a) Ordinary CP PL in chiral materials. The favored light handedness depends solely on the chirality of the luminescent material but independent of the emitting direction. (b)-(e) The ACPLE induced by finite current in chiral or inversion-symmetry-breaking achiral materials sandwiched by PN junctions. The favored light handedness depends on the emitting direction and also reverses with the direction of current $I$. For chiral materials the dirrection of ACPLE can be either parallel or perpendicular to the direction of current [(b)-(e)], but for inversion-symmetry-breaking achiral materials, the ACPLE can be only perpendicular to the current [(b)-(c)].
    }
    \label{fig1}
\end{figure*}

\section{Theory of ACPLE}

According to Fermi's golden rule, the optical emission rate of left- and right- handed CP light is determined by
\begin{equation}
    \begin{split}
        I_{L/R} = \frac{2\pi}{\hbar} \sum_{c,v,\bs{k}} |\langle v\bs{k}| H' |c\bs{k}\rangle |^2 f_{c\bs{k}} (1-f_{v\bs{k}}) \delta(\varepsilon_{c\bs{k}} - \varepsilon_{v\bs{k}} - \hbar\omega)
    \end{split}
\end{equation}
where $|c\bs{k}\rangle$ ($|v\bs{k}\rangle$) refers to the Bloch state of conduction (valance) band, $\varepsilon_{c\bs{k}}$ ($\varepsilon_{v\bs{k}}$) is the conduction (valance) band energy, $f_{c\bs{k}}$ is the electron occupation probability in the cunduction band and $1-f_{v\bs{k}}$ refers to the hole occupation in the valance band, and $H' = -e\bs{E} \cdot \bs{r} - \bs{m} \cdot \bs{B}$ is the light-matter interaction Hamiltonian with $e\bs{r}$ and $\bs{m}$ being the electric and magnetic moments, respectively, $\bs{E}$ and $\bs{B}$ being the electric and magnetic fields of light, respectively, and $\hbar\omega$ is the photon energy. For CP light with wave vector $\bs{q}$, the electric and magnetic fields are $\bs{E} = \frac{E_0}{\sqrt{2}} (\hat{\bs{e}}_1 \pm i \hat{\bs{e}}_2)$ and $\bs{B} = \frac{1}{c_0} \hat{\bs{q}} \times \bs{E}$ with $\hat{\bs{e}}_2$ and $\hat{\bs{e}}_2$ being the two unit vectors perpendicular to $\bs{q}$, $c_0$ is speed of light, and $\hat{\bs{q}} = \bs{q}/|\bs{q}|$ is the unit vector of $\bs{q}$. 

The difference of generated CP light emission can be calculated as
\begin{equation}
    \begin{split}\label{CPL}
        I_L &- I_R = -\frac{2\pi e^2}{\hbar} |E_0|^2 \sum_{c,v,\bs{k}}  f_{c\bs{k}} (1-f_{v\bs{k}}) \delta(\varepsilon_{c\bs{k}} - \varepsilon_{v\bs{k}} - \hbar\omega) \\ 
        &\left[ \frac{2}{ec_0} \textrm{Im} (r^{vc}_1 m^{cv}_1 + r^{vc}_2 m^{cv}_2) + \bs{\Omega}^{vc} \cdot \hat{\bs{q}} + O(m^2) \right],
    \end{split}
\end{equation}
where $r^{vc}_i =  \langle c\bs{k}| \bs{r} \cdot \hat{\bs{e}}_i | v\bs{k}\rangle$ and $m^{cv}_i = \langle c\bs{k}| \bs{m} \cdot \hat{\bs{e}}_i |v\bs{k}\rangle$ ($i=1,2$) refer to the electric and magnetic transition dipoles within the polarization plane of $\hat{\bs{e}}_1$ and $\hat{\bs{e}}_2$. $\bs{\Omega}^{vc} = -i \bs{r}^{vc} \times \bs{r}^{cv}$ is the interband Berry curvature \cite{Xu2018NatPhys,okyay2022second} whose expression is given by
\begin{equation}
    \Omega^{vc}_a = -i \epsilon_{abd} r^{vc}_b r^{cv}_d = -\hbar^2 \epsilon_{abd} \textrm{Im} \frac{\langle v\bs{k}| v_b |c\bs{k}\rangle \langle c\bs{k}| v_d | v\bs{k}\rangle }{(\varepsilon_{c\bs{k}} - \varepsilon_{v\bs{k}})^2}
\end{equation}
where $a,b,d=x,y,z$ refer to Cartesian corrdinates, $\epsilon_{abd}$ refers to the rank-3 antisymmetric Levi-Civita tensor and $v_b$, $v_d$ refer to the velocity operator. The first term on the second line of Eq. \eqref{CPL} gives rise to the ordinary CPLE or natural circular dichroism which has been extensively investigated in nonmagnetic chiral molecules and polymers \cite{Berova2000book, Longhi2016} as shown in Fig. \ref{fig1}(a). 

The second term in Eq.~\eqref{CPL} ($\bs{\Omega}^{vc} \cdot \hat{\bs{q}}$) gives rise to the ACPLE. This anomalous term is usually neglected before because the total Berry curvature vanishes in nonmagnetic materials such as organic molecules and polymers. It is not surprising that this term becomes significant in magnetic materials. Here, the magnetic order determines the sign of $\bs{\Omega}^{vc}$ and thereafter the sign of CPLE polarization. Furthermore, $\bs{\Omega}^{vc} \cdot \hat{\bs{q}}$ clearly indicates the emission-direction ($\hat{\bs{q}}$) dependence of CPLE, which was not awared before. Because the nonequilibrium current flow breaks time-reversal symmetry (TRS) even in nonmagnetic materials, the ACPLE term is also significant in CP EL. For a nonmagnetic system with TRS, the Berry curvature satisfies $\bs{\Omega}^{vc}(\bs{k}) = -\bs{\Omega}^{vc}(-\bs{k})$ which makes the anomalous term cancel with each other between $\bs{k}$ and $-\bs{k}$ at thermodynamic equilibrium. Therefore there are two cases the anomalous term can take effects: (i) The TRS is broken \textit{e.g.} by spontaneous magnetization $\bs{M}$ in magnetic materials (\textit{e.g.} CrI$_3$ monolayer); (ii) the system is in nonequilibrium state with the occupation probability satisfying $f_{c\bs{k}} \ne f_{c-\bs{k}}$ and $f_{v\bs{k}} \ne f_{v-\bs{k}}$ driving by an external electrical current \textit{i.e.} anomslous CP EL as shown in Fig. \ref{fig1}(b)-(c). Switching the current direction reverses the induced polarization. 
Particularly the anomalous term explicitly depends on the light-emitting direction $\hat{\bs{q}}$ resulting in the direction-dependent CP EL signals.

\textbf{Optical Berry curvature dipole.} In nonmagnetic materials, the system is driven out of thermal equilibrium by an external electrical dc current. The nonequilibrium distribution functions $f_{c\bs{k}}$ and $f_{v\bs{k}}$ can be determined by the Boltzmann equation under relaxation time approximation,
\begin{equation}
    \begin{split}\label{f_eh}
        f_{\alpha\bs{k}} = f^{0}_{\alpha\bs{k}} + \tau\frac{e}{\hbar} \bs{\mathcal{E}} \cdot \bs{\nabla_k} f^{0}_{\alpha\bs{k}} + O(\tau^2)
    \end{split}
\end{equation}
where $\alpha =c,v$ and $f^{0}_{\alpha\bs{k}} = \left[ e^{(\varepsilon_{\alpha\bs{k}}-\mu)/k_BT}+1 \right]^{-1}$ refers to the equilibrium Fermi distribution function, $\tau$ is the relaxation time, and $\bs{\mathcal{E}}$ is the external dc electric field. Substitute Eq. \eqref{f_eh} into Eq. \eqref{CPL}, the anomalous CP EL (second term of Eq. \eqref{CPL}) can be calculated as,
\begin{equation}
\begin{split}\label{ACPLE}
    I_L - I_R =& \frac{2\pi e^3 \tau}{\hbar^2} f_B(\hbar\omega) \mathcal{E}_a D_{ab}(\hbar \omega) \hat{q}_b + O(\tau^2), \\
    D_{ab}(\hbar \omega) =& \sum_{c,v,\bs{k}} (f^{0}_{v\bs{k}} - f^{0}_{c\bs{k}}) \partial_{k_a} \Omega^{vc}_b(\bs{k},\omega) \\
    \approx& \sum_{c,v,\bs{k}} \partial_{k_a} \Omega^{vc}_b(\bs{k},\omega)
\end{split}
\end{equation}
The identity $f^{0}_{c\bs{k}} (1 - f^{0}_{v\bs{k}}) = f_B(\varepsilon_{c\bs{k}} - \varepsilon_{v\bs{k}}) (f^{0}_{v\bs{k}} - f^{0}_{c\bs{k}})$ has been used in the derivation, and the last line of Eq. \eqref{ACPLE} is valid for large-gap semiconductors or insulators which are often used for optoelectronic materials. 
Here $f_B(\hbar\omega) = (e^{\hbar\omega/k_BT} - 1)^{-1}$ is the Bose distribution function. The optical Berry curvature refers to Berry curvature at the optical transition energy $\hbar\omega$,
\begin{equation}
    \Omega^{vc}_b(\bs{k},\omega)=\hat{\bs{e}}_b \cdot \bs{\Omega}^{vc}(\bs{k}) \delta(\varepsilon_{c\bs{k}} - \varepsilon_{v\bs{k}} - \hbar\omega)
\end{equation}
which represent the $b$-th component. The $D_{ab}(\hbar \omega)$, as expressed on the last line of Eq. \eqref{ACPLE}, represents the sum of the optical Berry curvature dipole (OBCD). It is a rank-2 pseudo tensor \textit{i.e.} it is even under TRS but odd under inversion, so that the anomalous CP EL in Eq. \eqref{ACPLE} can only occur in inversion-symmetry-breaking materials including chiral materials. 
In addition to the material-specific OBCD, the anomalous CP EL is also linear to the external driving field $\bs{\mathcal{E}}$ which rationalize previous experimental results about electrical switchable CP EL \cite{Zhang2014}. 

What was uncovered in previous experiments is that the CP EL also depends on the emission direction $\hat{\bs{q}}$. In optoelectronic literature, a dimensionless dissymmetry factor $g = (I_L - I_R)/\frac{1}{2}(I_L + I_R)$ is often used to describe the circular polarization ratio. The dissymmetry factor for the anomalous CP EL is 
\begin{equation}
\begin{split}\label{gEL}
    g_{EL} =& \frac{2e\tau}{\hbar} \frac{\mathcal{E}_a D_{ab} \hat{q}_b}{I_0} + O(\tau^2),
\end{split}
\end{equation}
where $I_0 = \sum_{c,v,\bs{k}} \left( |r^{vc}_1|^2 + |r^{vc}_2|^2 \right) \delta(\varepsilon_{c\bs{k}} - \varepsilon_{v\bs{k}} - \hbar\omega)$ is the ordinary absorption magnitude. According to Eq. \eqref{gEL}, the dissymmetry factor can be enhanced by improving the relaxation time $\tau$ and increasing electric field $\bs{\mathcal{E}}$ besides optimizing OBCD. 

In the context of the nonlienar anomalous Hall effect \cite{Sodemann2015}, the intra-band Berry curvature dipole was discussed and requires that the electric field $\mathcal{E}_a$ is orthogonal to $\Omega_b$, i.e., $a\neq b$.
In contrast, the ACPLE exists for both $a\neq b$ and $a = b$ in Eq.~\eqref{ACPLE}. If $a = b$, the material should be chiral because either inversion symmetry or mirror symmetry forces $D_{aa}$ to vanish. If $a \neq b$, the material only requires to break the inversion symmetry, which also applies for chiral materials. In the device, $a \neq b$ ($a = b$) means that light emission direction $\hat{q}_b$ can be normal [see Figs.\ref{fig1} (b) \& (c)] [parallel (see Figs. \ref{fig1} (d) \& (e))] to the current flow.

\section{Anomalous CP EL in inversion asymmetric materials}

Among the various inversion asymmetric materials found so far, the transition metal dichalcogenides (TMDs) are of particular interest because of the low-energy Dirac physics at Brillouin zone corners $K$ and $K'$ which exhibit valley-contrasting CP excitations processes. Zhang \textit{et al.} first demonstrated the electrically switchable CP EL in a WSe$_2$ monolayer transistor \cite{Zhang2014} and this effect was attributed to the triangular warping effect of the band structure \cite{Zhang2014, Wang2020}. Recently Pu \textit{et. al.} have experimentally demonstrated the room-temperature chiral light-emitting diode in strained WS$_2$ monolayer \cite{Pu2021WS2_exp}. In the following we illustrate the role of OBCD in the anomalous CP EL of strained WS$_2$ monolayer. 

\begin{figure}
    \centering
    \includegraphics[width=\linewidth]{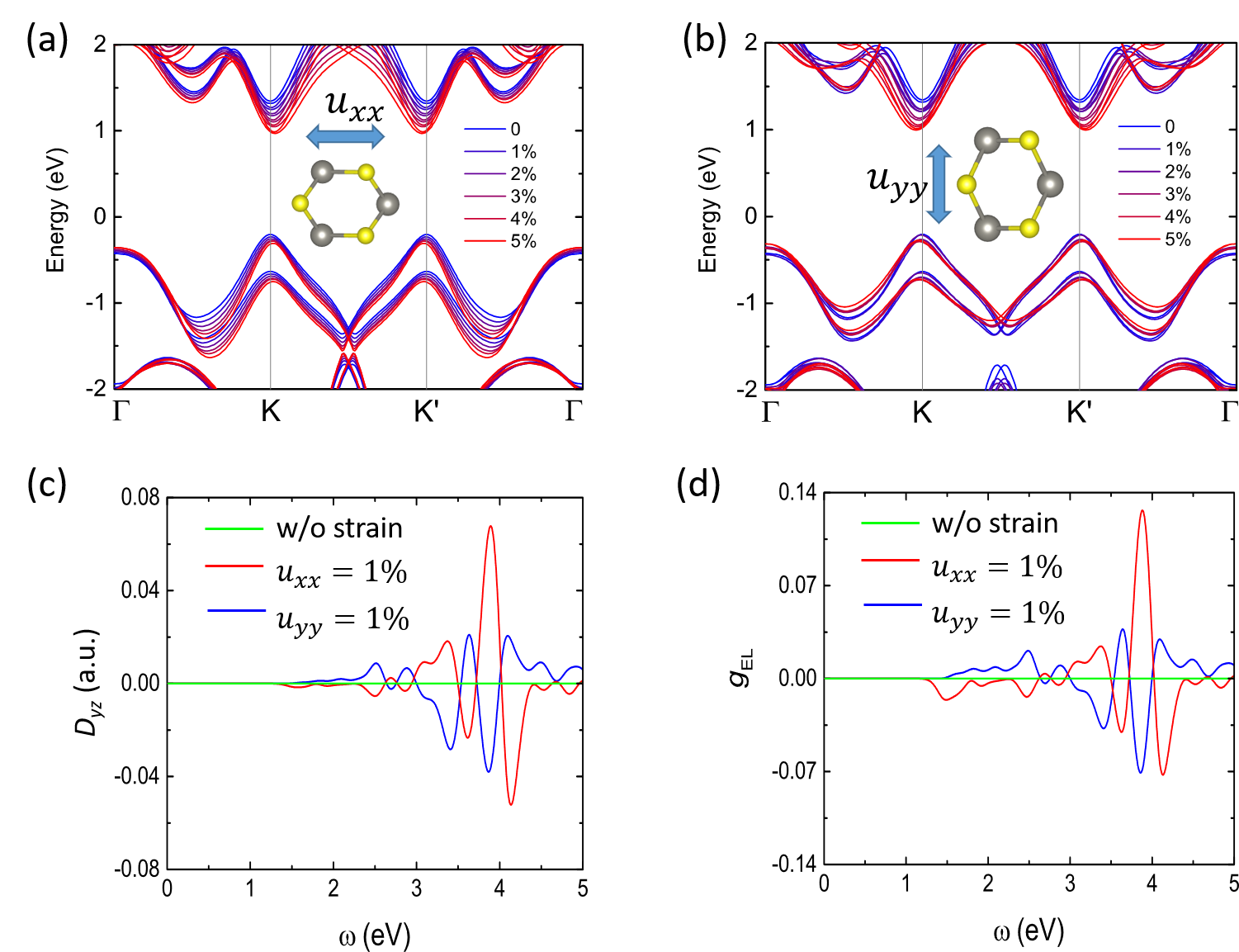}
    \caption{Anomalous CP EL in strained WS$_2$ monolayer. (a)-(b) Evolution of band structures of WS$_2$ monolayers under unidirectional strain $u_{xx}$ and $u_{yy}$, respectively. (c)-(d) Calculated OBCD and $g_{EL}$ for different strains. Under uniaxial strain $u_{xx}$ and $u_{yy}$, $D_{xz}(\omega)$ vanishes because of the combined mirror-time-reversal symmetry $M_y\mathcal{T}$. For $g_{EL}$, parameters $|\bs{\mathcal{E}}| = 10^4$ V/m and $\tau=6.6 $ fs are used in the calculations. 
    }
    \label{fig2}
\end{figure}

The crystal structure of unstrained WS$_2$ monolayer has a 3-fold rotation symmetry $C_{3z}$, under which the OBCD $D_{ab}$ ($a=x, y$, and $b=z$) vanishes for any $\omega$. Therefore any finite Berry dipole requires the breaking of $C_{3z}$ which can be realized by an externally applied uniaxial strain field $u_{xx}$ and $u_{yy}$. Figures \ref{fig2} shows the calculated band structure, OBCD, and $g_{EL}$ of WS$_2$ monolayer under different strain conditions based on first-principles methods (see Methods section for details). As is shown there, the calculated OBCD and $g_{EL}$ vanish for the unstrained case but become nonzero with finite $u_{xx}$ or $u_{yy}$ which is consistent with the symmetry analysis. 

On the other hand the strain-induced OBCD and $g_{EL}$ are of opposite sign between $u_{xx}$ and $u_{yy}$ as shown in Figs. \ref{fig2}(c)-(d) which enables strain engineering of CP EL in WS$_2$ monolayer. According to the semiclassical theory, the applied strain can induce an effective gauge potential \cite{Guinea2010NatPhys, Cazalilla2014PRL}
\begin{equation}\label{Aeff}
    \bs{A}^{\text{eff}} = \tau_z \frac{\beta}{a_0} (-2u_{xy}, u_{xx} - u_{yy})
\end{equation}
where $a_0$ is the lattice constant, $\beta$ is a parameter proportional to the electron-phonon coupling strength, and $u_{ab}$ ($a,b=x,y$) refers to the applied external strain tensor. $\tau_z=\pm 1$ is the valley index indicating $K$ and $K'$ valleys, respectively, reflecting the fact that the induced effective gauge potential are of opposite signs between two valleys, which is a manifestation of the TRS. According to Eq. \eqref{Aeff} the gauge potentials induced by different uniaxial strains $u_{xx}$ and $u_{yy}$ are of the opposite signs which moves the conduction and valance bands along opposite directions as is shown in Figs. \ref{fig2}(a)-(b). As a result the induced OBCD and $g_{EL}$ are of opposite sign for nonzero $u_{xx}$ and $u_{yy}$ as is shown in Figs. \ref{fig2}(c)-(d). The calculated value of $g_{EL}$ depends on magnetude of the applied field $\mathcal{E}$ and quality of sample (\textit{i.e.} relaxation time $\tau$). For moderate parameters $|\bs{\mathcal{E}}|=10^4$ V/m and $\tau=\hbar/\eta$ with $\eta=0.1$ eV the calculated maximum $g_{EL}$ is 0.1 at around 3.8 eV which is comparable to the experimental value of polarization rate between 9\% and 24\% \cite{Pu2021WS2_exp}. The $g_{EL}$ can be further enhanced by using larger electrical field or improving the sample quality according to Eq. \eqref{gEL}. However, it should be noted that $g_{EL}$ cannot become infinitely large with increasing $|\bs{\mathcal{E}}|$ and $\tau$ because it is bounded as $|g_{EL}|\le 2$ by its definition. This is because the relaxation time approximation in Eq. \eqref{f_eh} is valid only under moderate $|\bs{\mathcal{E}}|$ and $\tau$ which can be justified by the fact that the values of occupations $f_{\alpha\bs{k}}$ and $f^0_{\alpha\bs{k}}$ must be bounded by $0\le f_{\alpha\bs{k}}, f^0_{\alpha\bs{k}} \le 1$ so that $\tau \frac{e}{\hbar}\left| \bs{\mathcal{E}} \cdot \bs{\nabla_k} f^0_{\alpha\bs{k}} \right| = | f_{\alpha\bs{k}} - f^0_{\alpha\bs{k}} | \le 1$. For significantly large values of $|\bs{\mathcal{E}}|$ and $\tau$, the higher-order terms in $O(\tau^2)$ {and the achiral magnetic dipole contributions $O(m^2)$ of Eq. \eqref{CPL} must be included \cite{Tang2010PRL}}. The ACPLE in WS$_2$ monolayer exhibit rich switchable properties controlled by the direction of applied electrical current, strain, and emitting directions which is favorable for future optoelectronic device applications.

\section{Anomalous CP PL in magnetic materials}

CrI$_3$ is a van der Waals magnetic material with (anti-) ferromagnetic (interlayer) intralayer coupling \cite{Huang2017, Gibertini2019review}. Its monolayer exhibit in-plane ferromagnetism which breaks the TRS. The crystal structure has inversion symmetry which prohibit the ordinary term in Eq. \eqref{CPL} so that only the anomalous term contributes to the CP luminescence. 

We calculate the anomalous CP PL in the ferromangetic CrI$_3$ monolayer. The anomalous CP PL is determined by the OBC $\Omega(\bs{k}, \omega)$ which is the product of interband Berry curvature $\Omega^{vc}_{\bs{k}}$ and the joint density of states (JDOS) $g^{cv}_j(\bs{k},\omega) = \delta(\varepsilon_{c\bs{k}} - \varepsilon_{v\bs{k}} - \hbar\omega)$. Figure \ref{fig2}(a) shows the electronic band structure of CrI$_3$ monolayer along high-symmetry lines (HSLs) with a direct band gap of 0.85 eV at $\Gamma$ point. Derived from the band structure and Bloch wave functions along the HSLs, we plotted the optical excitation energy spectrum $\hbar\omega = \varepsilon_{c\bs{k}} - \varepsilon_{v\bs{k}}$ together with $\bs{k}$-resolved optical Berry curvature $\Omega(\bs{k},\omega) = \sum_{c,v} \Omega^{vc}_z(\bs{k},\omega)$ along high-symmetry lines (HSLs) in Fig. \ref{fig2}(b). The interband Berry curvature generally decreases with increasing $\omega$ because $\Omega^{cv}_{\bs{k}} \propto \frac{1}{(\varepsilon_{c\bs{k}} - \varepsilon_{v\bs{k}})^2} \propto \frac{1}{\omega^2}$, however, the JDOS $g^{cv}_j(\omega)$ generally increase with $\omega$ because of the increasing number of channels. As a result, there exist a maximum value of the optical Berry curvature $\Omega(\omega)$ as function of $\omega$ as is shown in the red line of Fig. \ref{fig2}(c). We also calculated the dissymmetry factor for the CP PL 
\begin{equation}\label{g_PL}
    g_{PL} = 2 \times \frac{I_L - I_R}{I_L + I_R} = \frac{2\Omega(\omega)}{I_0}
\end{equation}
as shown in the blue line of Fig. \ref{fig2}(c), where $\Omega(\omega) = \sum_{c,v,\bs{k}} \Omega^{vc}_z(\bs{k},\omega)$ refers to the total OBC. It should be noted that the anomalous CP PL in Eq. \eqref{g_PL} is determined only by band structure properties in contrast to the anomalous CP EL which also depends on relaxation time and external electrical field. Figure \ref{fig2}(d) shows the $\bs{k}$-resolved distribution of the optical Berry curvature $\Omega(\bs{k},\omega)$ at the peak value $\omega=1.14$ eV of the curve of optical Berry curvature $\Omega(\omega)$, which indicates the most prominent contribution from the highest valance band and the second lowest conduction band. The calculated maximum $g_{PL}$ of CrI$_3$ monolayer approaches about 0.42 at $\omega \approx 1.14$ eV which matches the order of magnitude with the experimental  $g\approx 1$ at $\omega=1.15$ eV \cite{Seyler2018}. Here, reversing the magnetization lead to opposite $\Omega(\omega)$ and thus opposite $g_{PL}$.

\begin{figure}
    \centering
    \includegraphics[width=\linewidth]{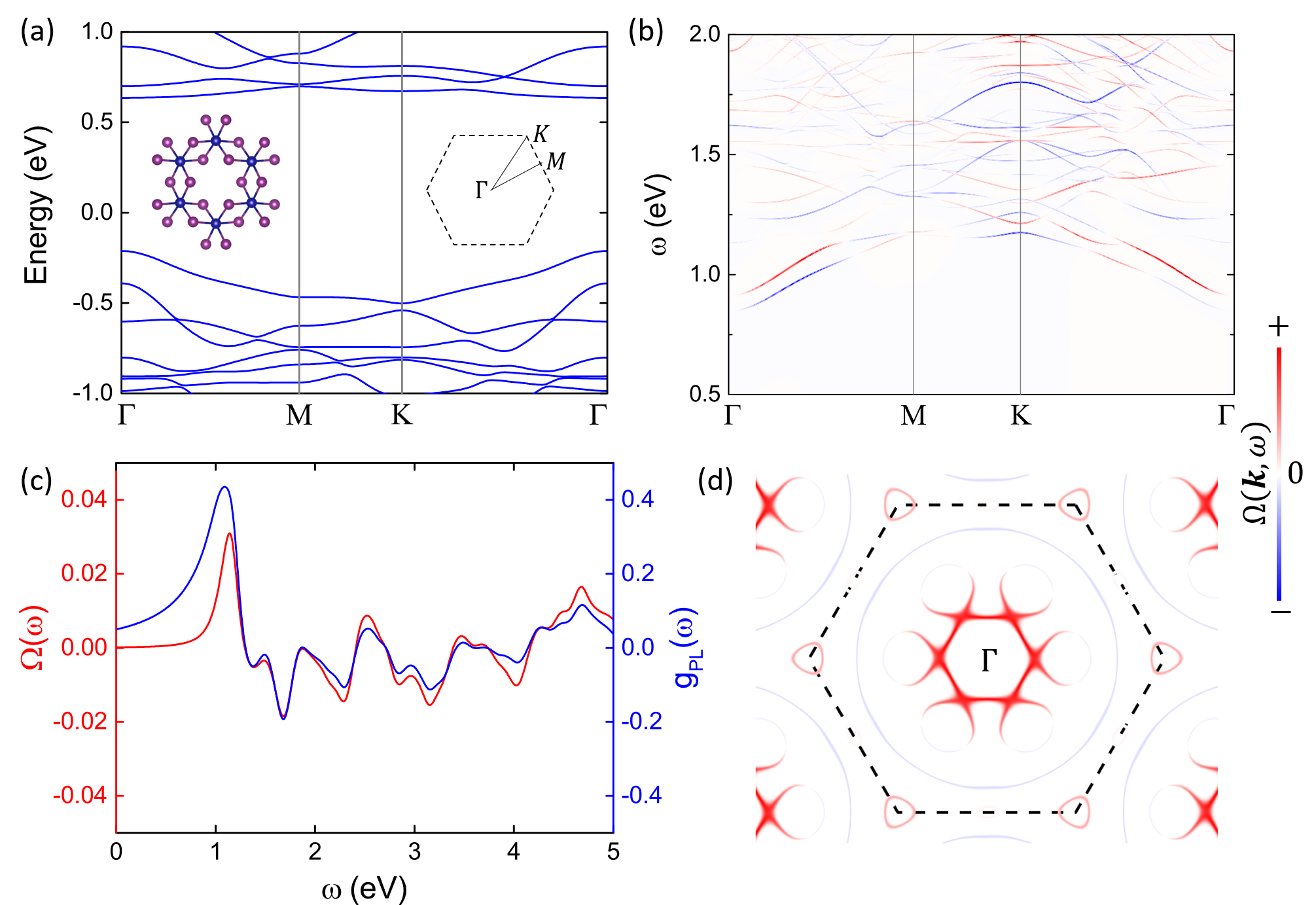}
    \caption{Anomslous CP PL in magnetic CrI$_3$ monolayer. (a) Band sturcture of CrI$_3$ monolayer whose crystal structure and Brillouin zone are shown in the insets. (b) Distribution of optical Berry curvature dipole $\Omega(\bs{k},\omega)$ along hight-symmetry lines. (c) Integrated total optical Berry curvature $\Omega(\omega) = \sum_{\bs{k}} \Omega(\bs{k},\omega) = \sum_{c,v,\bs{k}} \Omega^{vc}_{\bs{k}} \delta(\varepsilon_{c\bs{k}} - \varepsilon_{v\bs{k}} - \hbar\omega)$ and dissymmetry factor $g_{PL}=2(I_L - I_R)/(I_L + I_R)$ as functions of $\omega$. The peak value of $g_{PL} \sim 0.42$ at $\omega=1.14$ is comparable to the experimental value of maximum $g$ value of around 1 at $\omega = 1.15$ eV \cite{Seyler2018}. (d) Momentum resolved optical Berry curvature distribution $\Omega(\bs{k},\omega)$ at the peak value of $\omega = 1.14$ eV. }
    \label{fig3}
\end{figure}

\section{Discussion}

Because of the intimate relation between Berry curvature (dipole) with the linear (nonlinear) anomalous Hall effect (AHE) \cite{Xiao2010review, Nagaosa2010review, Sodemann2015}, the ACPLE satisfies a sum rule,
\begin{equation}
    \begin{split}
        \sigma^{\textrm{AH}} = \frac{1}{\pi} \int^{+\infty}_0 d\omega~ \frac{I_L - I_R}{f_B(\hbar\omega)}
    \end{split}
\end{equation}
where $\sigma^{\textrm{AH}}$ refer to the anomalous Hall conductivity in the plane perpendicular to the wave vector of light $\hat{\bs{q}}$. For 2D semiconductors/insulators, the linear AHE conductivity is quantized in multiples of $\frac{e^2}{h}$. Thus the integrated ACPLE spectrum give rise to topologically quantized signals which awaits for future experimental verification. 

The emission-direction dependence of ACPLE is a characteristic feature compared to the ordinary CPLE. In literature, CPLE is usually studied with the exciton-coupling model \cite{Swathi2020ChemCommun, Laidlaw2021ChemCommun} which cannot rationalize the current-direction and emission-direction dependence of ACPLE. 

\begin{figure}
    \centering
    \includegraphics[width=0.8\linewidth]{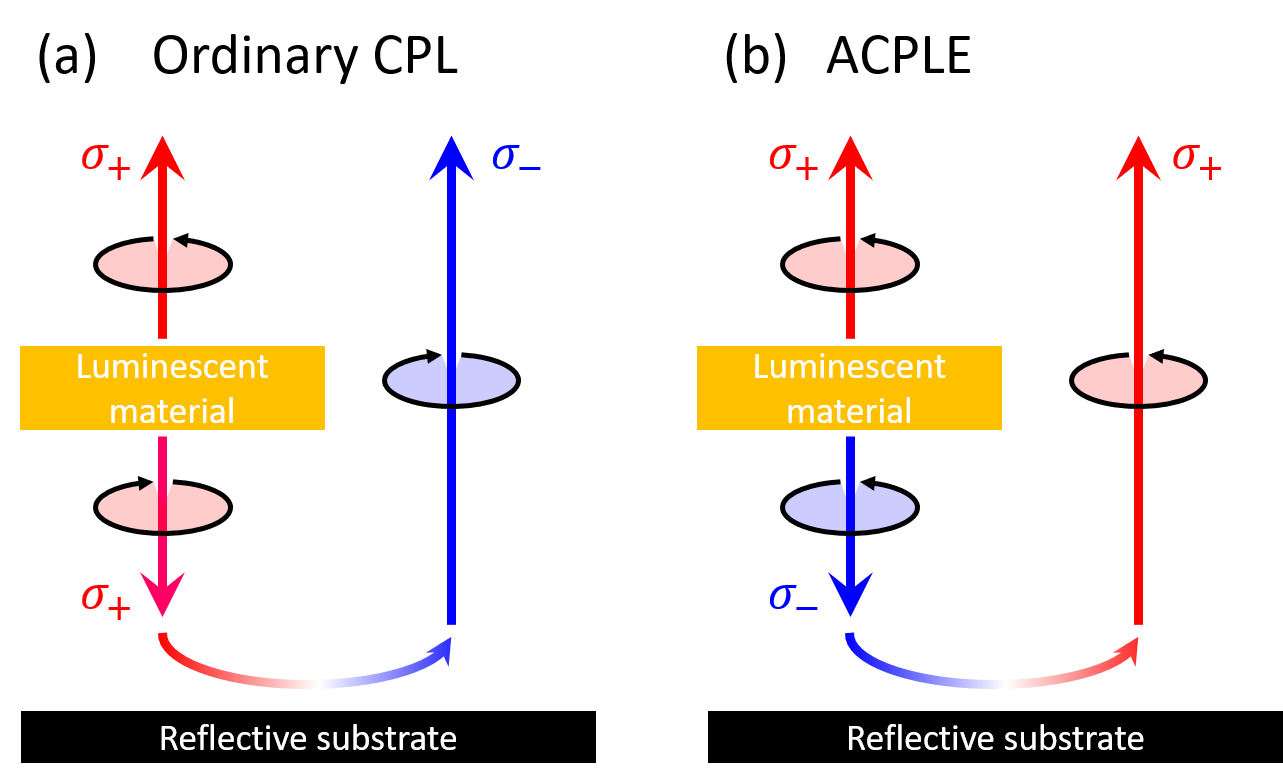}
    \caption{Comparison between (a) ordinary CPL and (b) ACPLE in the presence of reflective substrates. The net polarization of ordinary CPL is suppressed while it is enhanced for the ACPLE. 
    }
    \label{fig4}
\end{figure}

The emission-direction has a significant impact to enhance the polarization of a LED device. Figure \ref{fig4} shows the geometry of back-reflective LED. For the ordinary CPLE, the emitted CP light is of the same handedness for both directions. The handedness changed after back reflection(Fig. \ref{fig4}(a)). Therefore the net polarization diminishes when both light beams mix together. On the other hand, for the ACPLE, the handedness of the CP light for both directions are opposite. After reflection, both light beams exhibit the same handedness, avoiding the cancellation of polarization (Fig. \ref{fig4}(b)). Furthermore, the current-direction dependence enables high-speed switching of the light handedness by electric control, which essential for future optical quantum technology.

\section{Summary}
In summary, we proposed the quantum theory of anomalous circular polarized light emission based on the Berry curvature and Berry curvature dipole.
Our calculations on WS$_2$ and CrI$_3$ monolayers are consistent with recent experimental results. We propose that experiments can further detect the emission-direction dependence of the light handedness. Our findings deepen the understanding of band topology induced exotic optical phenomena and promise technology applications for high-speed, handedness-tunable optoelectrics. 

\section{Methods}

\textbf{First-principles calculations.} Our first-principles calculations are based on density functional theories as implemented in Vienna \textit{ab initio} Simulation Packages (VASP) \cite{Kresse1996}. The generalized gradient approximation of Perdew-Burke-Ernzerhof type \cite{Perdew1996PBE} has been used for the exchange-correlation functionals. The tight-binding Hamiltonians of CrI$_3$ and WS$_2$ monolayers are constructed using the code Wannier90 \cite{Mostofi2008, Mostofi2014, Pizzi2020}.

\begin{acknowledgments}
We thank Daniel Kaplan and Hengxin Tan for helpful discussions. B.Y. acknowledges the financial support by the European Research Council (ERC Consolidator Grant ``NonlinearTopo'', No. 815869) and  the Minerva foundation with funding from the Federal German Ministry for Education and Research.
\end{acknowledgments}


%

\end{document}